\documentclass[11pt]{article}

\usepackage[a4paper,margin=2.55cm,headheight=14pt]{geometry}
\usepackage{amsmath,amssymb}
\usepackage{graphicx}
\usepackage{booktabs}
\usepackage{microtype}
\usepackage{caption}
\usepackage{enumitem}
\usepackage{fancyhdr}
\usepackage{xcolor}
\usepackage{titlesec}
\usepackage[hidelinks]{hyperref}
\usepackage{authblk}
\newif\ifmarked \markedfalse
\newcommand{\rev}[1]{\ifmarked{\color{red}#1}\else#1\fi}
\newcommand{\revon}{\ifmarked\color{red}\fi}

\newcommand{\RBthick}{0.0618}
\newcommand{\TBthick}{0.8840}
\newcommand{\ABthick}{0.0542}
\newcommand{\WallBthick}{15.6}
\newcommand{\RBthin}{0.1194}
\newcommand{\TBthin}{0.8781}
\newcommand{\ABthin}{0.00255}
\newcommand{\WallBthin}{6.3}
\newcommand{\RExthick}{0.0620}
\newcommand{\TExthick}{0.8838}
\newcommand{\AExthick}{0.0542}
\newcommand{\RExthin}{0.1194}
\newcommand{\TExthin}{0.8781}
\newcommand{\AExthin}{0.00255}

\newcommand{\arXivI}{\href{https://arxiv.org/abs/2610.29336}{arXiv:2610.29336}}

\definecolor{sectiongray}{gray}{0.25}
\definecolor{rulegray}{gray}{0.72}

\setlist{itemsep=2pt,topsep=4pt}
\titleformat{\section}
  {\large\bfseries\color{sectiongray}}
  {\thesection}
  {0.65em}
  {}
  [\vspace{0.25ex}\color{rulegray}\titlerule]
\titleformat{\subsection}
  {\normalsize\bfseries}
  {\thesubsection}
  {0.55em}
  {}
\titlespacing*{\section}{0pt}{2.0ex plus .4ex minus .2ex}{1.0ex}
\titlespacing*{\subsection}{0pt}{1.5ex plus .3ex minus .2ex}{0.6ex}

\newcommand{\ud}{\mathrm{d}}
\newcommand{\MethodA}{FDTD--ADE}

\newcommand{\maybeincludegraphics}[2][]{\includegraphics[#1]{#2}}

\title{\vspace{-1.0cm}
  {\LARGE\bfseries Light pulse propagation through a dielectric slab.\\[0.2em]
  {\revon II.~Two roads to the same field}}\\[0.45em]
  {\large Numerical cross-validation of local (FDTD--ADE) and non-local
  (retarded integral-equation) solutions}
}
\author[1]{Giancarlo Reali\thanks{reali@unipv.it}}
\affil[1]{\small Dipartimento di Ingegneria Industriale e dell'Informazione,
Universit\`a di Pavia, via Ferrata 5a, 27100 Pavia, Italia}
\date{\small \rev{October} 2026}

\begin{document}
\maketitle
\thispagestyle{plain}

\begin{abstract}
\rev{Paper~I of this series~\cite{Reali2026pulse}} solves, via the retarded Green's functions, the local differential equations governing light propagation in a plane-parallel dielectric slab, obtaining a pair of retarded integral equations that represent, at the continuum level, the microscopic multiple-scattering picture underlying the local formulation. The two formulations are thus proven to be \emph{analytically} equivalent. The local system of differential equations is then solved numerically with a finite-difference time-domain scheme (FDTD--ADE). This report closes the loop by solving the retarded integral equations \emph{directly}, without passing through the local formulation, and comparing the result with the FDTD--ADE solution under the same physical and material conditions. The two independently implemented, structurally different numerical schemes agree within $0.02$--$1.3\%$ for the reported reflected and transmitted energy fractions; the absorbed fraction, computed independently in both schemes from the Lorentz damping term rather than merely inferred from $1-R-T$, agrees closely between the two methods, to the precision reported. \rev{Both solutions are also compared with an exact reference, the Airy formula weighted over the pulse spectrum: the direct solution reproduces it to within $0.3\%$, and its Richardson extrapolation to all digits shown.} The agreement persists for a few-cycle pulse and for a sub-wavelength slab, where naive fixed-point iteration of the integral equations is not a practical convergent strategy. A complementary set of movie-style, nine-frame visualizations is also presented, illustrating the same equivalence directly in the field rather than only in integrated observables. To the author's knowledge, a direct numerical cross-validation of these two formulations in this genuinely time-dependent regime has not previously been reported.
\end{abstract}

\section{Introduction}
\label{sec:intro}

\rev{Paper~I~\cite{Reali2026pulse}} establishes, by direct substitution of the retarded Green's functions, that the integral equations
\begin{align}
E(z,t) &= E_{\mathrm{in}}(z,t) - \frac{2\pi}{c}\int_0^L \! \ud z' \,
\frac{\partial P\!\left(z',t-|z-z'|/c\right)}{\partial t},
\label{eq:IE-field}\\
\frac{\partial P(z,t)}{\partial t} &= \frac{Ne^2}{m}\int_0^t \! \ud t' \,
 e^{-\gamma(t-t')}\frac{\rev{w}\cos\!\left(\rev{w}(t-t')\right)-\gamma\sin\!\left(\rev{w}(t-t')\right)}{\rev{w}}
 E(z,t'),
\label{eq:IE-pol}
\end{align}
are the formal solution of the local system
\begin{equation}
\frac{\partial^2 E}{\partial z^2}-\frac{1}{c^2}\frac{\partial^2 E}{\partial t^2}
=\frac{4\pi}{c^2}\frac{\partial^2 P}{\partial t^2},
\qquad
\ddot P + 2\gamma\dot P + \omega_0^2 P = \kappa E,
\qquad \kappa\equiv\frac{Ne^2}{m}.
\label{eq:ML}
\end{equation}
\rev{with $w=(\omega_0^2-\gamma^2)^{1/2}$. Equations~\eqref{eq:IE-field} and \eqref{eq:IE-pol} are Eqs.~(4) and (8) of Paper~I, and Eqs.~\eqref{eq:ML} are its Eqs.~(1) and (5); the notation and the parameters (Table~\ref{tab:params}) are the same in the two papers.}
The analytic proof of equivalence is conclusive on its own. The purpose of the present report is different and complementary: to demonstrate, by two independent numerical implementations, that the two formulations lead to the same numerical solution within the resolution of the calculations.

It is worth being explicit about what makes this comparison non-trivial, beyond bookkeeping. Equations~\eqref{eq:ML} are a \emph{field-centric, local} description: the medium is a continuum whose polarization responds, at each point $z$, only to the field \emph{at that same point} and \emph{at that same instant}, through an ordinary differential equation in time alone, while the field itself is propagated forward by a hyperbolic partial differential equation whose finite characteristic speed $c$ builds causality in implicitly, through the structure of the differential operators -- retardation is never written down explicitly anywhere in Eqs.~\eqref{eq:ML}. Equations~\eqref{eq:IE-field}--\eqref{eq:IE-pol} are instead a \emph{source-centric, non-local} description, in the spirit of the microscopic multiple-scattering formalism developed in the context of quantum collision theory \cite{GoldbergerWatson} and previously adapted, at a pedagogical level, to continuous light propagation through a dielectric slab \cite{Reali2014}: the slab is a continuum of infinitesimal polarizable constituents, each obeying the same local oscillator equation in isolation, but each also radiating, via the explicit retarded Green's function $|z-z'|/c$, into every other constituent's driving field, including, in the limit $z'\to z$, its own. Reflection and refraction are not written into a local material response; they \emph{emerge} from the coherent sum, over the whole slab, of these causally-retarded contributions. Because the interaction is retarded, this picture retains the finite propagation time between different spatial locations, which is particularly relevant for ultrashort pulses, for which retardation cannot be represented solely by a steady-state phase factor.

This is, in miniature, a very old fork in theoretical physics: whether to describe a causal
system through a local equation of motion, or through a non-local, retarded-propagator
integral. Section~\rev{\ref{sec:conclusions}} returns to this point once the numerical evidence is in hand.

Both descriptions are exact and, since Eqs.~\eqref{eq:ML} and \eqref{eq:IE-field}--\eqref{eq:IE-pol} are analytically identical, both must produce the same field. What is not automatic is that two independent numerical schemes, built around these two different mathematical structures -- one marching a local PDE forward in time with explicit spatial derivatives, the other solving a retarded convolution with no spatial derivative operator at all -- will actually agree once implemented, discretized, and run. That numerical cross-check is the subject of this report.

The comparison should be understood at three distinct levels: (i) the \emph{analytic equivalence} of the local and retarded formulations, established in \rev{Paper~I}; (ii) the \emph{numerical equivalence} of independently discretized solutions, which is the principal subject here; and (iii) the agreement of derived observables such as $R$, $T$, the absorbed fraction, and the envelope delay. Section~\ref{sec:results} addresses levels (ii) and (iii) with integrated quantities and full space-time fields; Sec.~\ref{sec:movie} adds a movie-style, frame-by-frame view of the same equivalence for a single representative case.

\section{Method A: FDTD--ADE}
\label{sec:methodA}

Method A follows the scheme of \rev{Paper~I (its Sec.~3)}: Eq.~\eqref{eq:ML} is discretized on a uniform space-time grid with a second-order leapfrog update for $E$ and an auxiliary-differential-equation (ADE) recursion for $P$, together with a total-field/scattered-field (TF/SF) source injection and first-order Mur absorbing boundaries. ADE formulations are standard tools for incorporating frequency-dependent material response into FDTD simulations \cite{Taflove,Taflove2,Prokopidis}. The material, pulse, and slab parameters are kept identical between the two numerical approaches (see Table \ref{tab:params}); the numerical time stepping is allowed to differ because the two algorithms have different requirements.

\begin{table}[!htbp]
\centering
\small
\begin{tabular}{@{}ll@{}}
\toprule
Quantity & Value \\
\midrule
Carrier wavelength $\lambda_c$ & 800 nm \\
\rev{Pulse FWHM, field envelope} & 5.9 fs (2.2 cycles) \\
\rev{Pulse FWHM, intensity} & \rev{4.2 fs ($\approx1.6$ cycles)} \\
UV resonance wavelength $\lambda_0$ & 120 nm \\
Quality factor $Q=\omega_0/(2\gamma)$ & 50 \\
Target refractive index $n(\omega_c)$ & 1.45 \\
Group index $n_g(\omega_c)$ & 1.4675 \\
Thick slab thickness & $3\,\mu\mathrm{m}$ (round trip $2nL/c=29.0\,\mathrm{fs}$) \\
Thin slab thickness & $150\,\mathrm{nm}$ (round trip $2nL/c=1.45\,\mathrm{fs}$) \\
\bottomrule
\end{tabular}
\caption{Material and pulse parameters used in both calculations. The physical parameters are identical; the numerical time step is method-dependent.}
\label{tab:params}
\end{table}

\section{Method B: direct solution of the retarded integral equations}
\label{sec:methodB}

\subsection{A first attempt: the Neumann/Born series}
\label{sec:divergence}

Because Eqs.~\eqref{eq:IE-field}--\eqref{eq:IE-pol} are linear, a natural first approach is
a Neumann (multiple-scattering, or Born) series,
\begin{equation}
E=\sum_{n\ge0}\Delta E^{(n)},
\qquad
\Delta E^{(0)}=E_{\mathrm{in}},
\qquad
\Delta E^{(n+1)}=\mathcal{K}\big[\Delta E^{(n)}\big].
\label{eq:neumann}
\end{equation}
The integral operator $\mathcal{K}$ acts in two steps, the same two steps used to build Method B in Sec.~\ref{sec:directscheme}: Eq.~\eqref{eq:IE-pol} turns $\Delta E^{(n)}$ into the polarization it drives, and Eq.~\eqref{eq:IE-field} turns that polarization into the next-order field $\Delta E^{(n+1)}$. Diagrammatically, order $n$ is therefore light that has scattered $n$ times inside the slab: $\Delta E^{(1)}$ is the single-scattering contribution, $\Delta E^{(2)}$ the double-scattering contribution, and so on.

\begin{figure}[!htbp]
\centering
\maybeincludegraphics[width=0.70\textwidth]{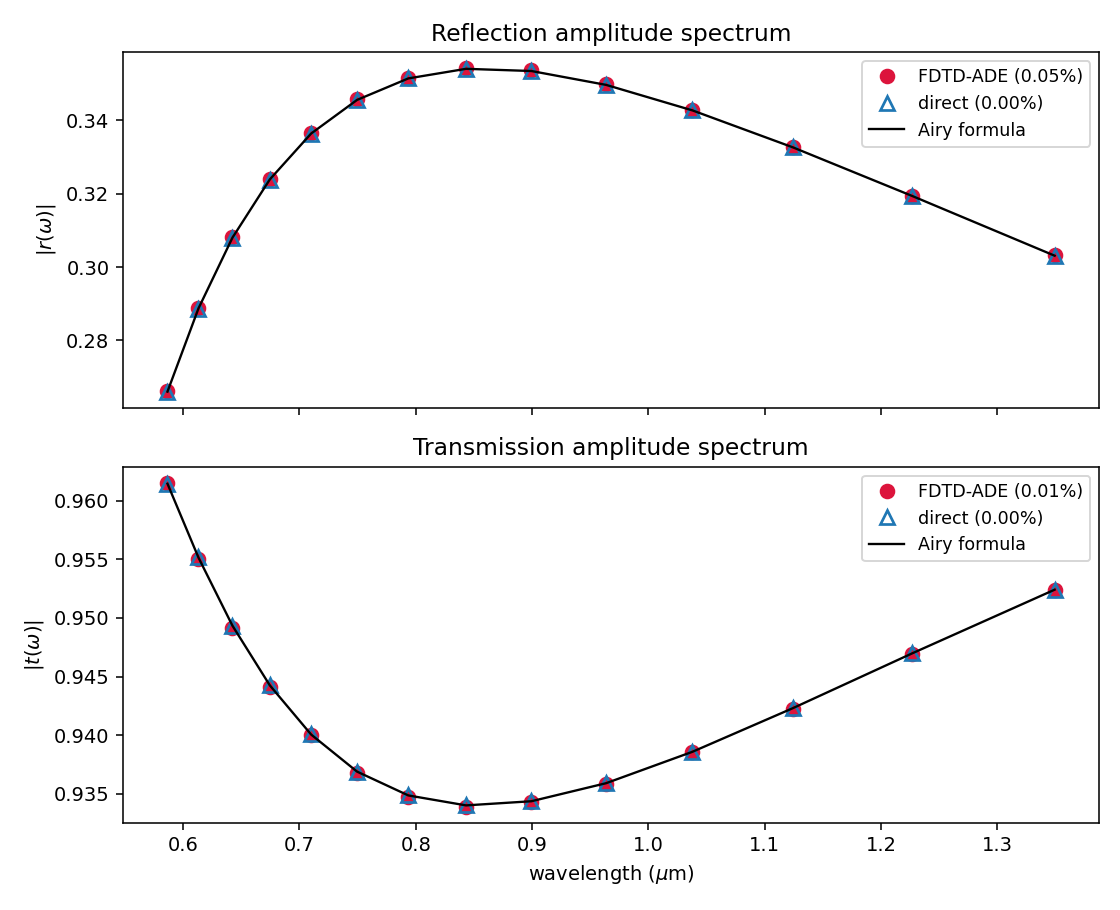}
\caption{Reflection and transmission spectra extracted from Method A (filled circles) and Method B (open triangles), compared with the analytic Airy prediction (solid line), evaluated with the complex, frequency-dependent Lorentz index.}
\label{fig:5}
\end{figure}

To test whether this expansion is a good approximation here, we evaluate the first-order term at the carrier frequency $\omega_c(=2\pi c/\lambda_c)$, at the front interface $z=0$ where the reflected field is read off\rev{, for the thin slab ($L=150$~nm)}. In frequency domain, with $\tilde E(z,\omega)$ and $\tilde P(z,\omega)$ denoting the Fourier transforms in time of $E(z,t)$ and $P(z,t)$ \rev{(time dependence $e^{-i\omega t}$)}, Eq.~\eqref{eq:IE-pol} becomes, pointwise in $z$, the algebraic relation $\tilde P(z,\omega)=\chi(\omega)\tilde E(z,\omega)$ with the susceptibility
\begin{equation}
\chi(\omega)=\frac{n^2-1}{4\pi}=\frac{\kappa}{\omega_0^2-\omega^2-2i\gamma\omega},
\label{eq:chi}
\end{equation}
where the first equality is the standard relation to the complex refractive index already used in the frequency-domain Airy comparison of Sec.~\ref{sec:results} (Fig.~\ref{fig:5}), and the second is its explicit form for the single-resonance Lorentz model of Eq.~\eqref{eq:ML}. Applying Eq.~\eqref{eq:IE-field} to the resulting polarization and integrating over the slab thickness $L$ gives
\begin{equation}
\left.\frac{\Delta \tilde E^{(1)}(0,\omega)}{\tilde E_{\mathrm{in}}(0,\omega)}\right|_{\omega=\omega_c}
=
\pi\,\chi(\omega_c)\left[e^{\rev{2i\omega_cL/c}}-1\right]
\approx 0.5.
\label{eq:first-order}
\end{equation}
This is already a poor sign: a legitimate perturbative correction should be small compared with unity, not half of it. Physically, this reflects the fact that a sub-wavelength slab is not a dilute medium from the point of view of the constituents' mutual coupling: every point of the slab is within a small fraction of an optical cycle's retardation of every other point, so the ``single-scattering'' contribution already sums coherently over the whole slab rather than over a thin, weakly-coupled layer.

The numerical experiment confirms that this is not merely a slowly-converging series: the partial sums of Eq.~\eqref{eq:neumann} -- equivalently, the Picard iterates of the fixed-point map $E\mapsto E_{\mathrm{in}}+\mathcal{K}[E]$ -- grow without bound rather than converge to the correct solution. For a linear fixed-point map, convergence for arbitrary initial data is governed by the spectral radius \rev{$\rho$} of $\mathcal{K}$, which must be below unity. \rev{Since $\mathcal K$ is time-invariant, the time-domain iteration acts on each frequency component separately, and the relevant quantity is the spectral radius $\rho(\omega)$ of the frequency-domain operator $(\mathcal K\tilde E)(z)=(2\pi i\omega/c)\,\chi(\omega)\int_0^L e^{i\omega|z-z'|/c}\,\tilde E(z')\,\ud z'$, maximized over all frequencies. Discretizing this operator on the slab and computing its eigenvalues gives, for the thin slab, $\rho=0.62$ at the carrier frequency, so that a monochromatic wave at $\omega_c$ alone would still yield a convergent series; but $\rho$ exceeds unity above about $1.7\,\omega_c$, where the incident spectrum is below $10^{-6}$ of its peak, and reaches $\rho\approx10^2$ near the UV resonance, where $|\chi|$ is about $Q$ times larger than at the carrier. For the thick slab $\rho\approx4.6$ already at $\omega_c$. Frequency components with $\rho>1$, however weak in the incident pulse or seeded only by round-off, are amplified at every iteration, and the series diverges, as observed. The size of the first-order term \eqref{eq:first-order} alone is therefore an indication, not the cause, of the divergence.} This is a property of the iterative Born expansion, not of the underlying linear integral equations \eqref{eq:IE-field}--\eqref{eq:IE-pol}, which remain well posed and are solved directly, without any series expansion, in Sec.~\ref{sec:directscheme}. Convergence conditions of this kind are classical in scattering theory \cite{ColtonKress}; \rev{operator-norm bounds of this kind, for the closely related Born series of diffuse waves, are given in}~\cite{MoskowSchotland}.

\subsection{A direct time-marching scheme}
\label{sec:directscheme}

We instead solve Eqs.~\eqref{eq:IE-field}--\eqref{eq:IE-pol} by direct time marching on a uniform spatial grid $\{z_i\}_{i=0}^{N_z-1}$ spanning $[0,L]$, with $z_0=0$ and $z_{N_z-1}=L$. Two numerical issues require special treatment: the zero-delay self-term and the retarded coupling between distinct grid points. The approach is conceptually related to time-marching methods for retarded potential integral equations \cite{Davies1994,Sauter2013}; Laguerre-transform/boundary-element approaches provide another established route for retarded kernels \cite{Chapko2019}.

\paragraph{Spatial discretization and quadrature.}
The slab is discretized into $N_z$ uniformly spaced points,
\begin{equation}
z_i=i\,\Delta z,
\qquad
\Delta z=\frac{L}{N_z-1},
\qquad
 i=0,1,\ldots,N_z-1.
\label{eq:grid}
\end{equation}
The spatial integral in Eq.~\eqref{eq:IE-field} is approximated by the trapezoidal rule, with weights
\begin{equation}
w_i=\Delta z \quad (1\le i\le N_z-2),
\qquad
w_0=w_{N_z-1}=\frac{\Delta z}{2}.
\label{eq:weights}
\end{equation}
Time is discretized as $t_n=n\Delta t$.

\paragraph{(i) Zero-delay self-interaction.}
At $z_j=z_i$, the retardation in Eq.~\eqref{eq:IE-field} vanishes. The corresponding discrete contribution therefore couples the current-time field $E_i^n$ to the current-time polarization derivative $\dot P_i^n$, making the update implicit. To avoid a nonlinear or iterative inner solve, the ADE recursion associated with Eq.~\eqref{eq:ML} is combined algebraically with the zero-delay term.

Using the central difference
\begin{equation}
\dot P_i^n=\frac{P_i^{n+1}-P_i^{n-1}}{2\Delta t},
\label{eq:Pdot}
\end{equation}
we obtain a linear relation
\begin{equation}
\dot P_i^n=\beta E_i^n+\dot P_{i,\mathrm{hist}}^n,
\qquad
\beta=\frac{\kappa}{2\Delta t\left(\frac{1}{\Delta t^2}+\frac{\gamma}{\Delta t}\right)},
\label{eq:selfrel}
\end{equation}
where the history term is known from previously computed polarization values,
\begin{equation}
\dot P_{i,\mathrm{hist}}^n=
\frac{1}{2\Delta t}
\left[
\frac{-\omega_0^2P_i^n+\frac{2P_i^n}{\Delta t^2}
+P_i^{n-1}\left(\frac{\gamma}{\Delta t}-\frac{1}{\Delta t^2}\right)}
{\frac{1}{\Delta t^2}+\frac{\gamma}{\Delta t}}
-P_i^{n-1}
\right].
\label{eq:Phist}
\end{equation}
Substitution into the discretized field equation gives the non-iterative update
\begin{equation}
E_i^n=
\frac{\mathrm{known}_i^n-\frac{2\pi}{c}w_i\dot P_{i,\mathrm{hist}}^n}
{1+\frac{2\pi}{c}w_i\beta},
\label{eq:explicit}
\end{equation}
where
\begin{equation}
\mathrm{known}_i^n=E_{\mathrm{in}}(z_i,t_n)
-\frac{2\pi}{c}\sum_{j\ne i}w_j\dot P\!\left(z_j,t_n-\frac{|z_i-z_j|}{c}\right).
\label{eq:known-explicit}
\end{equation}
The self-term is thus eliminated algebraically rather than through a fixed-point iteration. This removes the specific divergence mechanism encountered in the naive Picard iteration. It does not, by itself, constitute a proof of unconditional stability for arbitrary material parameters or discretizations.

\paragraph{(ii) Exact retarded lookup in time.}
For $j\ne i$, the retardation is strictly positive. In general it need not coincide with a time-grid point and would therefore require temporal interpolation. We choose
\begin{equation}
S\equiv\frac{c\Delta t}{\Delta z}=0.5,
\end{equation}
so that
\begin{equation}
\frac{|z_i-z_j|}{c}=2|i-j|\Delta t.
\label{eq:delay}
\end{equation}
Every pairwise retardation is consequently an integer multiple of $\Delta t$. The retarded values can therefore be retrieved by array lookup, with no temporal interpolation error. This does \emph{not} eliminate spatial quadrature, time-discretization, or self-term errors. We precompute the integer delay matrix
\begin{equation}
D_{ij}=2|i-j|,
\label{eq:delaymatrix}
\end{equation}
and access the history as $\dot P_j^{\,n-D_{ij}}$. If $n-D_{ij}<0$, the retarded contribution is set to zero, enforcing causality.

The resulting algorithm is non-iterative in the spatial coupling at each time step. Its stability and convergence should nevertheless be understood as properties of the complete discretized scheme, rather than inferred solely from the algebraic elimination of the self-term.

The reflected and transmitted fields are evaluated at the slab boundaries according to the definitions used in \rev{Paper~I},
\begin{equation}
E_r(0,t)=E(z_0,t)-E_{\mathrm{in}}(z_0,t),
\qquad
E_t(L,t)=E(z_{N_z-1},t).
\label{eq:RTfields}
\end{equation}

For the movie-style visualization of Sec.~\ref{sec:movie}, the field at any observation point $z_{\mathrm{obs}}$ outside $[0,L]$ -- where the incident, reflected, and transmitted wavepackets are seen approaching, splitting and separating -- is obtained from the same already-computed polarization history via a single evaluation of
\begin{equation}
E(z_{\mathrm{obs}},t) = E_{\mathrm{in}}(z_{\mathrm{obs}},t) - \frac{2\pi}{c}\sum_j w_j\,\dot P\!\left(z_j,\,t-\frac{|z_{\mathrm{obs}}-z_j|}{c}\right),
\label{eq:external}
\end{equation}
with retarded values obtained by linear interpolation of the stored history -- a post-processing step requiring no additional time-marching.

\subsection{Grid convergence and computational cost}
\label{sec:convergence-test}

In Method A, the time step is chosen from the standard one-dimensional FDTD Courant relation, with $S=0.4$. In Method B, $S=0.5$ is selected to make the retarded delays exact integer multiples of the time step. Thus, the physical conditions are identical, but the two numerical discretizations are not identical.

The wall time is the elapsed computer time required for a complete simulation. Method A has a local field update and scales approximately as $O(N_z)$ per time step. The direct retarded summation in Method B scales as $O(N_z^2)$ per time step in the straightforward implementation, explaining its substantially larger execution time.

The convergence tables\rev{, Tables~\ref{tab:convB_thin} and~\ref{tab:convB_thick},} report $R$, $T$, and the absorption fraction inferred from the balance
\begin{equation}
A_{\mathrm{inf}}\equiv1-R-T.
\label{eq:Ainf}
\end{equation}
Because $A_{\mathrm{inf}}$ is defined from $R$ and $T$, the identity $R+T+A_{\mathrm{inf}}=1$ is not, by itself, an independent energy-conservation test; \rev{the tables therefore also report the absorbed fraction $A_{\mathrm{indep}}$ computed independently, from the work done by the damping term (Appendix~\ref{app:energy}), which coincides with $A_{\mathrm{inf}}$ to all digits shown. The tables are written directly by the program of Method~B.}

\begin{table}[!htbp]
\centering
\small
{\revon
\begin{tabular}{@{}lrrrrrr@{}}
\toprule
$N_z$ & $\Delta t$ (as) & $R$ & $T$ & $A_{\mathrm{inf}}$ & $A_{\mathrm{indep}}$ & wall time (s) \\
\midrule
21 & 12.51 & 0.11901 & 0.87843 & 0.00256 & 0.00256 & 0.6 \\
41 & 6.25 & 0.11928 & 0.87817 & 0.00255 & 0.00255 & 0.8 \\
81 & 3.13 & 0.11934 & 0.87810 & 0.00255 & 0.00255 & 1.5 \\
161 & 1.56 & 0.11936 & 0.87809 & 0.00255 & 0.00255 & 6.3 \\
\bottomrule
\end{tabular}
}
\caption{Grid convergence of Method B for the thin slab. The inferred absorption fraction changes very little over the tested range, while $R$ and $T$ approach stable values. The final column gives the measured wall time\rev{; the runs are executed in parallel, one per core}.}
\label{tab:convB_thin}
\end{table}

\begin{table}[!htbp]
\centering
\small
{\revon
\begin{tabular}{@{}lrrrrrr@{}}
\toprule
$N_z$ & $\Delta t$ (as) & $R$ & $T$ & $A_{\mathrm{inf}}$ & $A_{\mathrm{indep}}$ & wall time (s) \\
\midrule
101 & 50.03 & 0.05892 & 0.88659 & 0.05449 & 0.05449 & 0.8 \\
201 & 25.02 & 0.06120 & 0.88452 & 0.05428 & 0.05428 & 2.5 \\
401 & 12.51 & 0.06177 & 0.88400 & 0.05423 & 0.05423 & 15.6 \\
601 & 8.34 & 0.06188 & 0.88390 & 0.05422 & 0.05422 & 79.6 \\
\bottomrule
\end{tabular}
}
\caption{Grid convergence of Method B for the thick slab. The inferred absorption fraction is substantially larger than for the thin slab. Between the two finest grids, $R$ and $T$ change by \rev{less than about $0.2\%$ and $0.01\%$, respectively}, supporting the use of $N_z=401$ for the production comparison.}
\label{tab:convB_thick}
\end{table}

The convergence data also allow an empirical order-of-convergence estimate for the successively halved grids. Applying the standard three-level Richardson estimate to the first three \rev{resolutions gives second-order behavior for both slabs: $p\approx2.0$ for $R$ and for $T$, for the thin as for the thick slab. The corresponding Richardson extrapolations, $R=0.11937$, $T=0.87808$ (thin) and $R=0.06196$, $T=0.88383$ (thick), coincide to all digits shown with the exact values of Sec.~\ref{sec:exact}.} These values should be regarded as empirical diagnostics rather than formal proofs of the asymptotic order of the complete method.

\section{Physical expectations and numerical results}
\label{sec:results}

Before examining the numerical results, it is useful to state the main physical expectations. For a single interface with $n=1.45$, the Fresnel reflection coefficient is
\begin{equation}
R_F=\left(\frac{n-1}{n+1}\right)^2\approx0.034.
\end{equation}
For a slab, multiple internal reflections interfere and are described in the monochromatic limit by the Airy formula \cite{BornWolf}. For the thick slab, the round-trip time ($29.0$ fs) is much longer than the pulse duration ($5.9$ fs), so the front-surface reflection and the back-surface echo are temporally separated. For the thin slab, the round-trip time ($1.45$ fs) is shorter than one optical cycle, so multiple reflections overlap and interfere coherently, modifying the net reflected pulse.

A few-cycle pulse in a dispersive Lorentz medium can also exhibit a precursor-like feature ahead of the main pulse. The main envelope is expected, near the carrier frequency and for moderate pulse distortion, to propagate at the group velocity
\begin{equation}
v_g=\frac{c}{n_g}.
\end{equation}
Any such feature seen in the calculations would be consistent with the causal high-frequency response of the Lorentz model, but a definitive identification as a Sommerfeld--Brillouin precursor would require a dedicated asymptotic or spectral analysis \cite{Oughstun2007,Oughstun2013}, which has not been performed here. A separate check, reported in Sec.~\ref{sec:movie}, further shows that the small leading tail visible under strong magnification ahead of the idealized incident pulse is a mathematical property of the infinite-duration Gaussian source used to drive both simulations, not by itself evidence of medium-generated dispersive forerunners; searching the leading and trailing edges of the transmitted pulse and of the back-surface echo for a distinct, medium-generated secondary lobe found none resolvable above the discretization noise for this material and pulse bandwidth. This is a negative result at the resolution tested, not a rigorous exclusion.

\subsection{Reflected and transmitted fields: thick slab}

Method A (\MethodA) is plotted with solid lines and Method B (direct integral equations) with dashed lines. The contrasting line styles make small discrepancies visible even when the curves nearly overlap.

\begin{figure}[!htbp]
\centering
\maybeincludegraphics[width=0.72\textwidth]{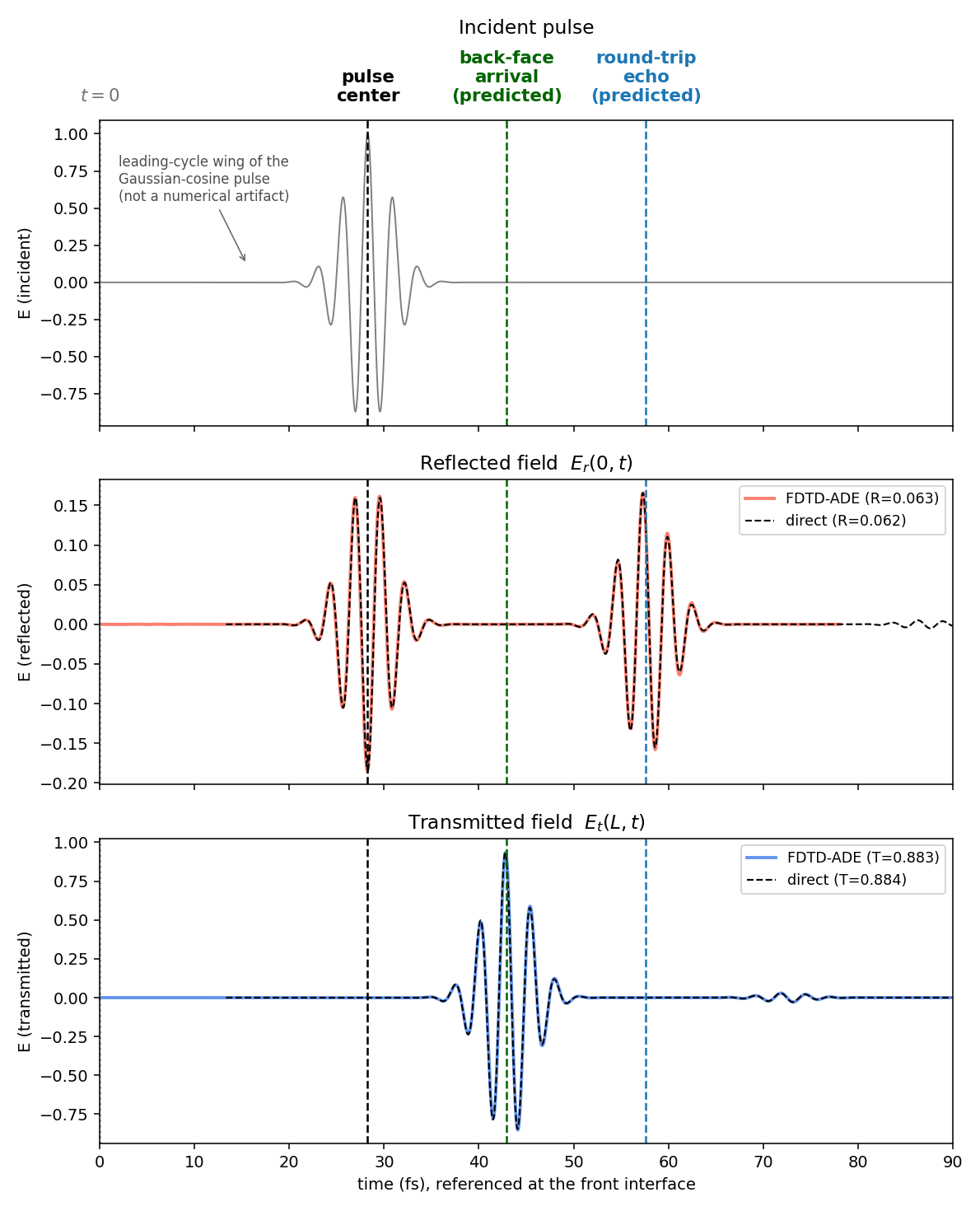}
\caption{Thick slab ($L=3\,\mu\mathrm{m}$): incident, reflected, and transmitted fields for Method A (solid) and Method B (dashed). The predicted-arrival markers (pulse center, back-face group-delay arrival, and round-trip echo) apply to both calculations.}
\label{fig:1}
\end{figure}

Figure~\ref{fig:1} shows the direct-integral result overlaid on the FDTD--ADE calculation. At the plotted scale the curves are visually indistinguishable. The integrated energy fractions are $R=0.0626$ (A) versus $0.0618$ (B), a relative difference of about $1.3\%$, and $T=0.8826$ (A) versus $0.8840$ (B), a relative difference of about $0.16\%$\rev{; the exact values of Sec.~\ref{sec:exact} are $R=\RExthick$ and $T=\TExthick$}.

\subsection{Thin and thick slabs: reflected field}

\begin{figure}[!htbp]
\centering
\maybeincludegraphics[width=0.72\textwidth]{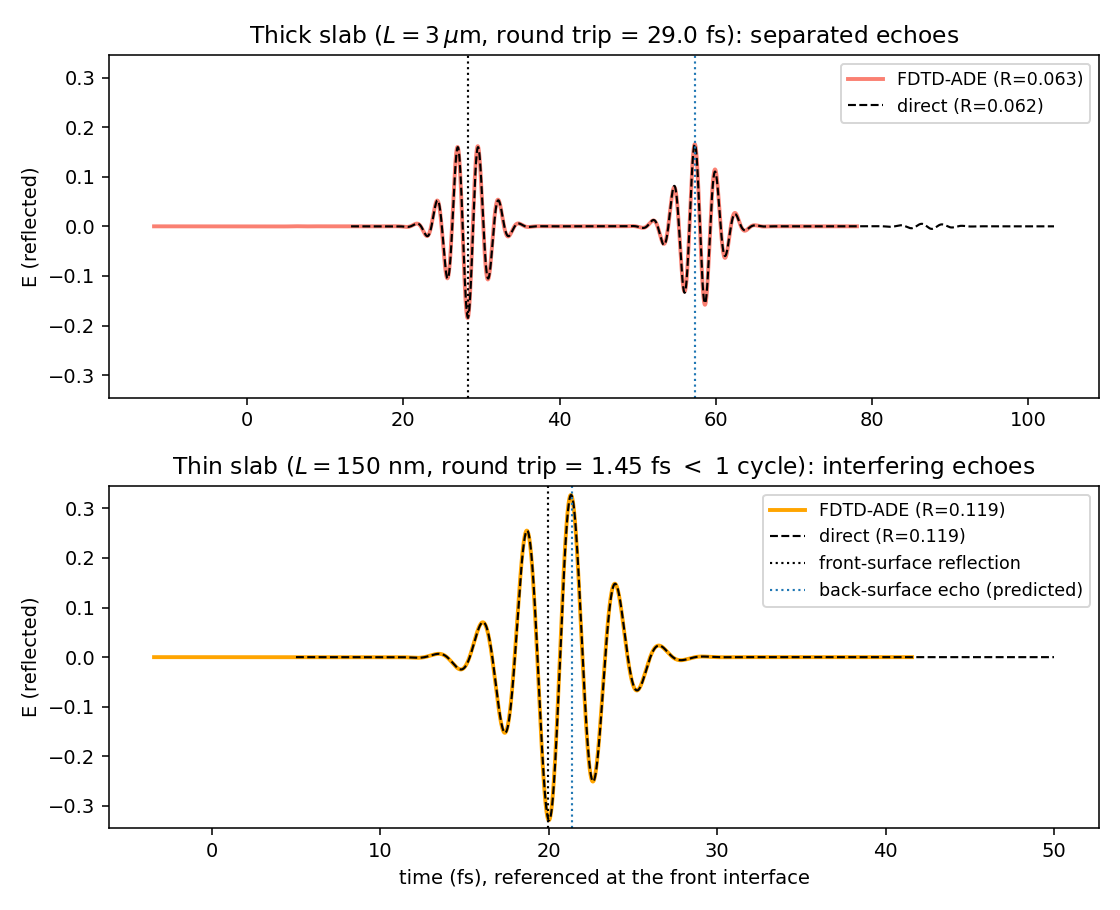}
\caption{Reflected field for both slab thicknesses and both numerical methods. Top: thick slab, where the echoes are temporally separated. Bottom: thin slab, where multiple reflections overlap within the pulse duration.}
\label{fig:2}
\end{figure}

\rev{Figure~\ref{fig:2} compares the reflected fields of the two slabs.} The thin-slab case provides the more demanding test because the internally reflected contributions overlap strongly. Nevertheless, the reflected energy fractions are $R=0.1195$ (A) and $R=0.1194$ (B), differing by only about $0.084\%$.

\subsection{Space-time and envelope maps}

\begin{figure}[!htbp]
\centering
\maybeincludegraphics[width=\textwidth]{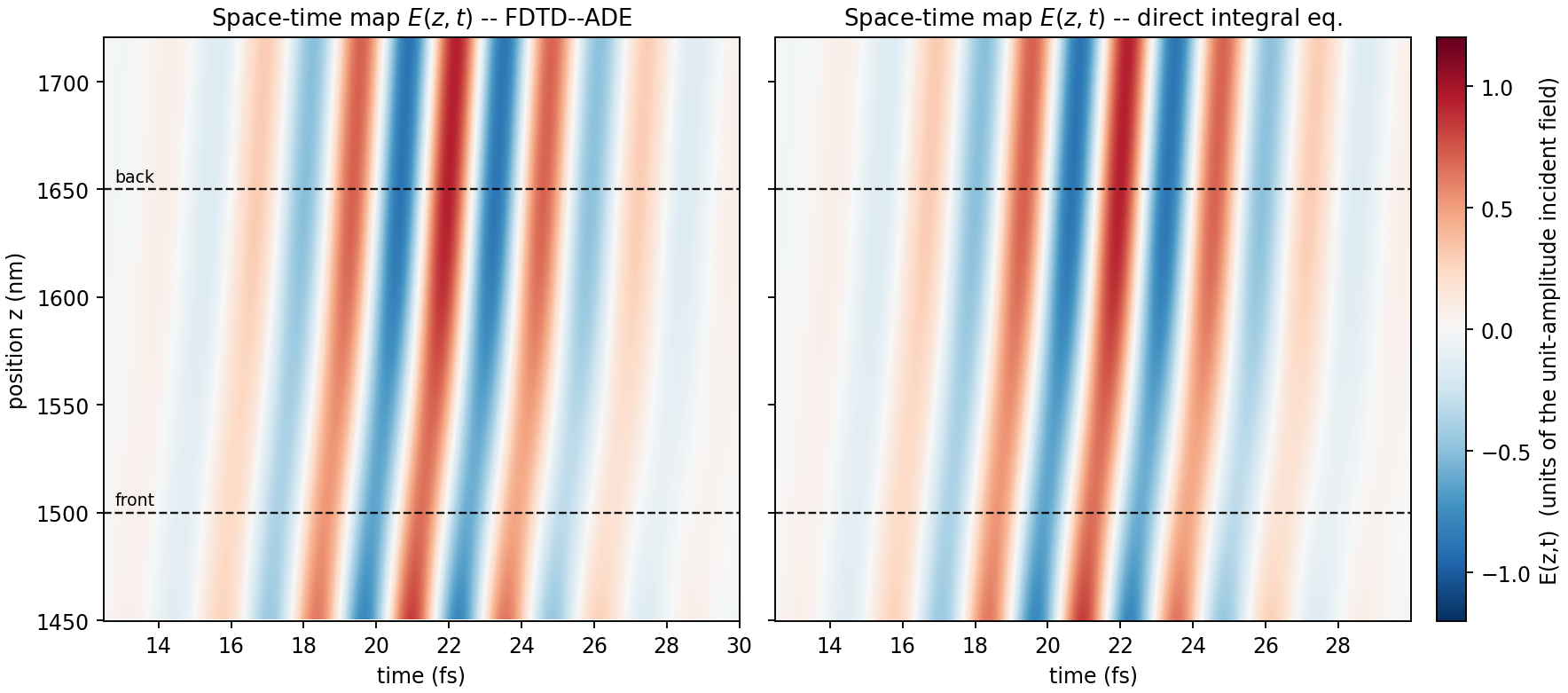}
\caption{Space-time map of $E(z,t)$ for the thin slab: Method A (left) and Method B (right). The dashed horizontal lines mark the slab interfaces. The internal fringe pattern is reproduced in both calculations.}
\label{fig:3}
\end{figure}

Figure~\ref{fig:3} provides a direct visualization of the field inside the slab. The internal pattern results from the coherent superposition of incident and multiply reflected contributions. Interpolating Method A's map onto Method B's native grid over the common domain and time window gives an RMS field difference of $0.0036$, on a field normalized so the incident peak has unit amplitude -- a quantitative field-norm error consistent with the close visual agreement.

\begin{figure}[!htbp]
\centering
\maybeincludegraphics[width=\textwidth]{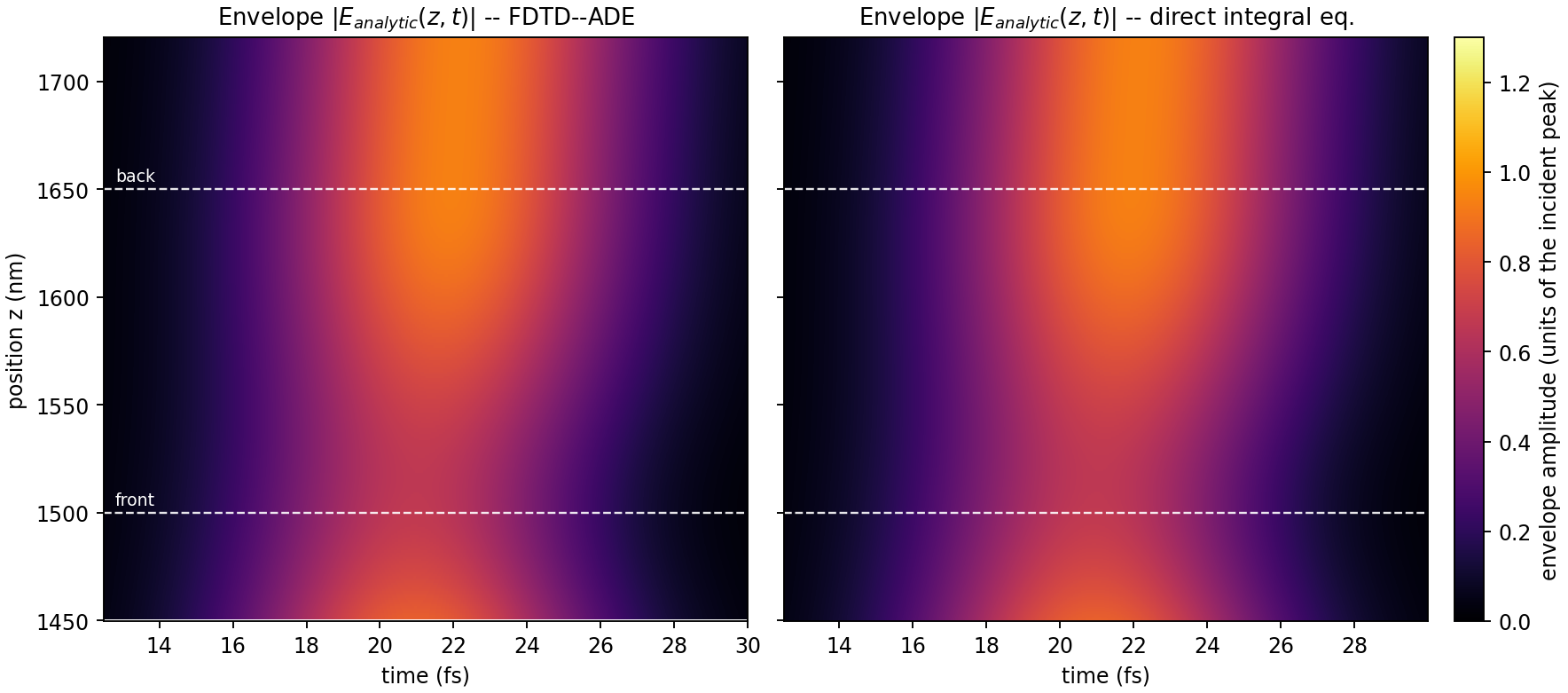}
\caption{Envelope maps $|E_{\mathrm{analytic}}(z,t)|$ for the thin-slab simulations. The two methods reproduce the same dominant envelope structure and interference region.}
\label{fig:4}
\end{figure}

\paragraph{Envelope propagation and group velocity.}
The measured transit time is defined precisely as
\begin{equation}
\tau = t_{\mathrm{peak}}(L) - t_{\mathrm{peak}}(0),
\qquad
t_{\mathrm{peak}}(z) \equiv \arg\max_t \big|E_{\mathrm{analytic}}(z,t)\big|,
\label{eq:tpeak}
\end{equation}
the arrival time of the peak of the Hilbert-transform envelope at each interface -- the same envelope estimator already used, and shown, throughout Secs.~\ref{sec:results}--\ref{sec:movie}. Group velocity then follows, by definition rather than by a second, independent measurement, as $v_g^{\mathrm{meas}}=L/\tau$; Table~\ref{tab:energy}'s $v_g$ row is accordingly a direct restatement of the transit times quoted below, not an independent check on them. Because the pulse contains only a few optical cycles, this envelope-peak estimator is itself an approximation, most accurate for moderate pulse reshaping; the comparison with the predicted $n_gL/c$ is therefore interpreted as a consistency check rather than as a high-precision determination of $v_g$.

\subsection{Frequency-domain Airy consistency}

The reflection and transmission spectra are obtained by Fourier transforming the time-domain fields, and both numerical results closely follow the plotted Airy prediction, computed with the same frequency-dependent complex refractive index used in both simulations, over the displayed spectral range.

\subsection{Energy fractions, group-delay check, and computational cost}

\begin{table}[!htbp]
\centering
\small
\rev{\begin{tabular}{@{}lccc|ccc@{}}
\toprule
& \multicolumn{3}{c|}{Thick slab ($L=3\,\mu\mathrm{m}$)}
& \multicolumn{3}{c}{Thin slab ($L=150\,\mathrm{nm}$)} \\
& Method A & Method B & exact & Method A & Method B & exact \\
\midrule
$R$ & 0.0626 & \RBthick & \RExthick & 0.1195 & \RBthin & \RExthin \\
$T$ & 0.8826 & \TBthick & \TExthick & 0.878 & \TBthin & \TExthin \\
$A_{\mathrm{inf}}=1-R-T$ & 0.0548 & \ABthick & \AExthick & 0.0026 & \ABthin & \AExthin \\
$A_{\mathrm{indep}}$ (dissipation integral) & 0.0542 & \ABthick & --- & 0.0026 & \ABthin & --- \\
\midrule
$v_g=L/\tau$ ($10^8$ m/s) & 2.06 & 2.07 & --- & --- & --- & --- \\
Wall time (s) & 0.12 & \WallBthick & --- & 0.19 & \WallBthin & --- \\
\bottomrule
\end{tabular}}
\caption{Summary of energy fractions, envelope-velocity measurements, and execution times, each shown to three significant figures where the underlying run supports it. $A_{\mathrm{indep}}$ is computed independently, from the volume- and time-integrated dissipation-rate density $(2\gamma/\kappa)(\partial P/\partial t)^2$ of the driven, damped oscillator, rather than inferred from $1-R-T$; it agrees with $A_{\mathrm{inf}}$ to the precision shown \rev{for Method~B and for the thin slab of Method~A, and within about $1\%$ for the thick slab of Method~A ($0.0542$ against $0.0548$, a difference of the order of the discretization error of that run),} corroborating the $R+T+A=1$ balance as a genuine physical check rather than a tautology. The thin-slab Method~A entries are limited to two significant figures by the precision of the run reported here; Method~B's are taken from the finest grid of Table~\ref{tab:convB_thin}\rev{, and the thick-slab ones from the row $N_z=401$ of Table~\ref{tab:convB_thick}; the exact values are those of Eq.~\eqref{eq:exact}}. $v_g$ is obtained as $L/\tau$ from the transit times $\tau$ quoted in the text (Eq.~\eqref{eq:tpeak}), not from a second, independent measurement.}
\label{tab:energy}
\end{table}

The maximum relative discrepancy among the reported $R$ and $T$ values is about $1.3\%$, occurring for the thick-slab reflectance. For the thin slab, the corresponding discrepancies are below $0.1\%$. The execution times show the expected computational trade-off: for these benchmark runs, the direct retarded calculation is \rev{one to two orders of magnitude} slower than FDTD--ADE.

The predicted one-way group delay through the thick slab is
\begin{equation}
\tau_g=\frac{n_gL}{c}=14.685\,\mathrm{fs}.
\end{equation}
The measured transit times are $14.560$ fs for Method A ($0.85\%$ below the prediction) and $14.522$ fs for Method B ($1.11\%$ below the prediction). These values are consistent with the expected group delay at the accuracy resolved by the present calculation.

\subsection{\rev{An exact reference for the energy fractions}}
\label{sec:exact}
{\revon For a linear, time-invariant system the energy fractions of a pulse follow exactly from the monochromatic Airy coefficients $r(\omega)$ and $t(\omega)$ of the slab, evaluated with the complex Lorentz index $\tilde n(\omega)$, weighted by the incident spectrum. By Parseval's theorem,
\begin{equation}
R=\frac{\int_0^\infty |r(\omega)|^2\,|\tilde E_{\mathrm{in}}(\omega)|^2\,\ud\omega}{\int_0^\infty |\tilde E_{\mathrm{in}}(\omega)|^2\,\ud\omega},\qquad
T=\frac{\int_0^\infty |t(\omega)|^2\,|\tilde E_{\mathrm{in}}(\omega)|^2\,\ud\omega}{\int_0^\infty |\tilde E_{\mathrm{in}}(\omega)|^2\,\ud\omega},\qquad
A=1-R-T.
\label{eq:exact}
\end{equation}
These expressions involve no discretization in $z$ or $t$; the integrals are evaluated on a fine frequency grid, and they provide the independent benchmark against which both numerical methods can be judged. They give $R=\RExthick$, $T=\TExthick$, $A=\AExthick$ for the thick slab and $R=\RExthin$, $T=\TExthin$, $A=\AExthin$ for the thin slab (Table~\ref{tab:energy}). Method~B reproduces the thin-slab values to all digits shown and the thick-slab values to within $0.3\%$ at $N_z=401$; the Richardson extrapolations of Sec.~\ref{sec:convergence-test} coincide with them. The thick-slab reflectance of Method~A is $1.0\%$ above the exact value, consistent with the excess of the front-surface reflectivity found in Paper~I at the same grid spacing ($\Delta z=15$~nm). Most of the $1.3\%$ difference between the two methods on this quantity is therefore a discretization effect of the FDTD run, not a difference between the formulations.\par}

\section{A complementary, movie-style visualization}
\label{sec:movie}

\begin{figure}[!htbp]
\centering
\maybeincludegraphics[width=0.95\textwidth]{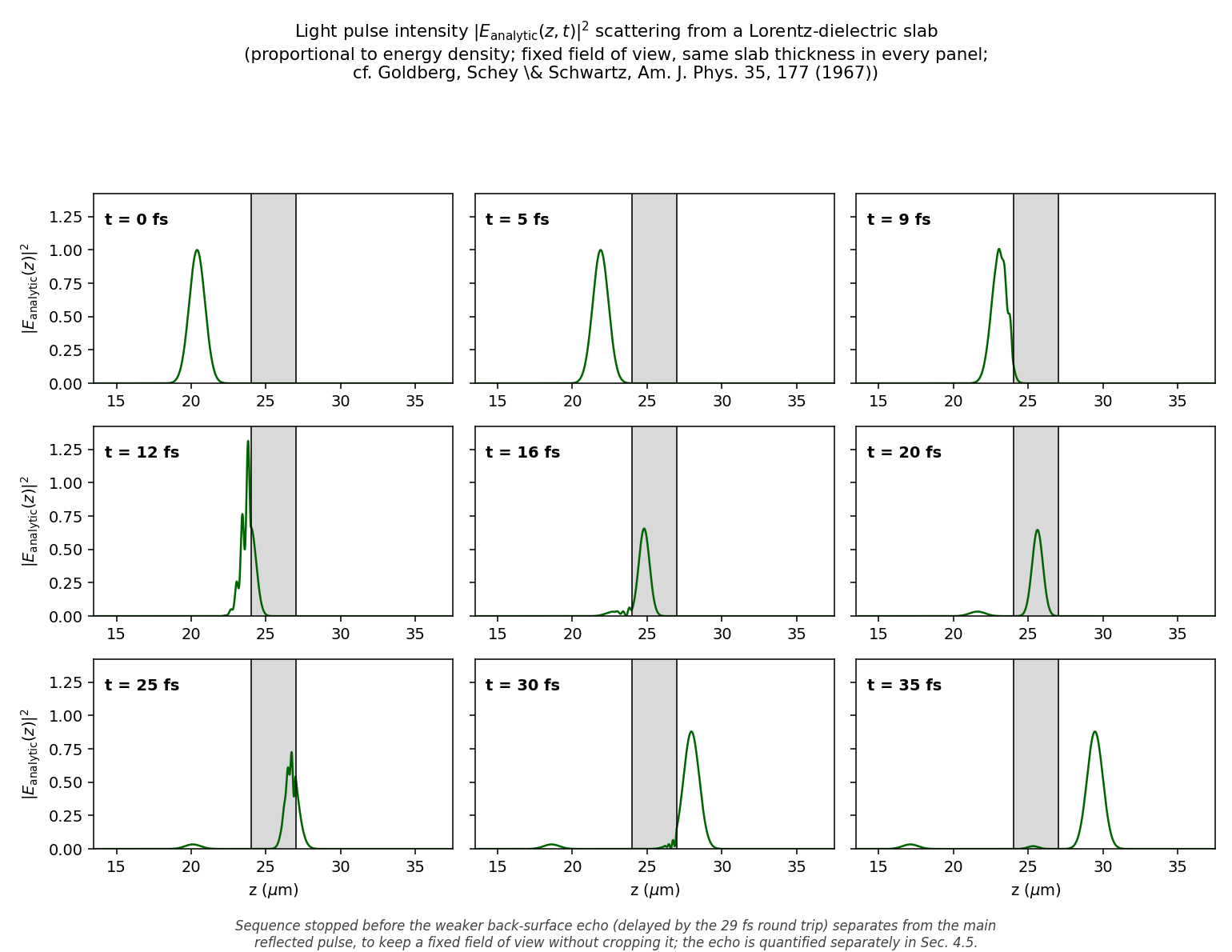}
\caption{FDTD--ADE: intensity $|E_{\mathrm{analytic}}(z,t)|^2$, proportional to energy density -- the direct analog of $|\psi(x,t)|^2$ in the reference wave-packet movies -- at nine snapshot times spanning approach, interaction, and separation.}
\label{fig:movie_intensity_fdtd}
\end{figure}

\begin{figure}[!htbp]
\centering
\maybeincludegraphics[width=0.95\textwidth]{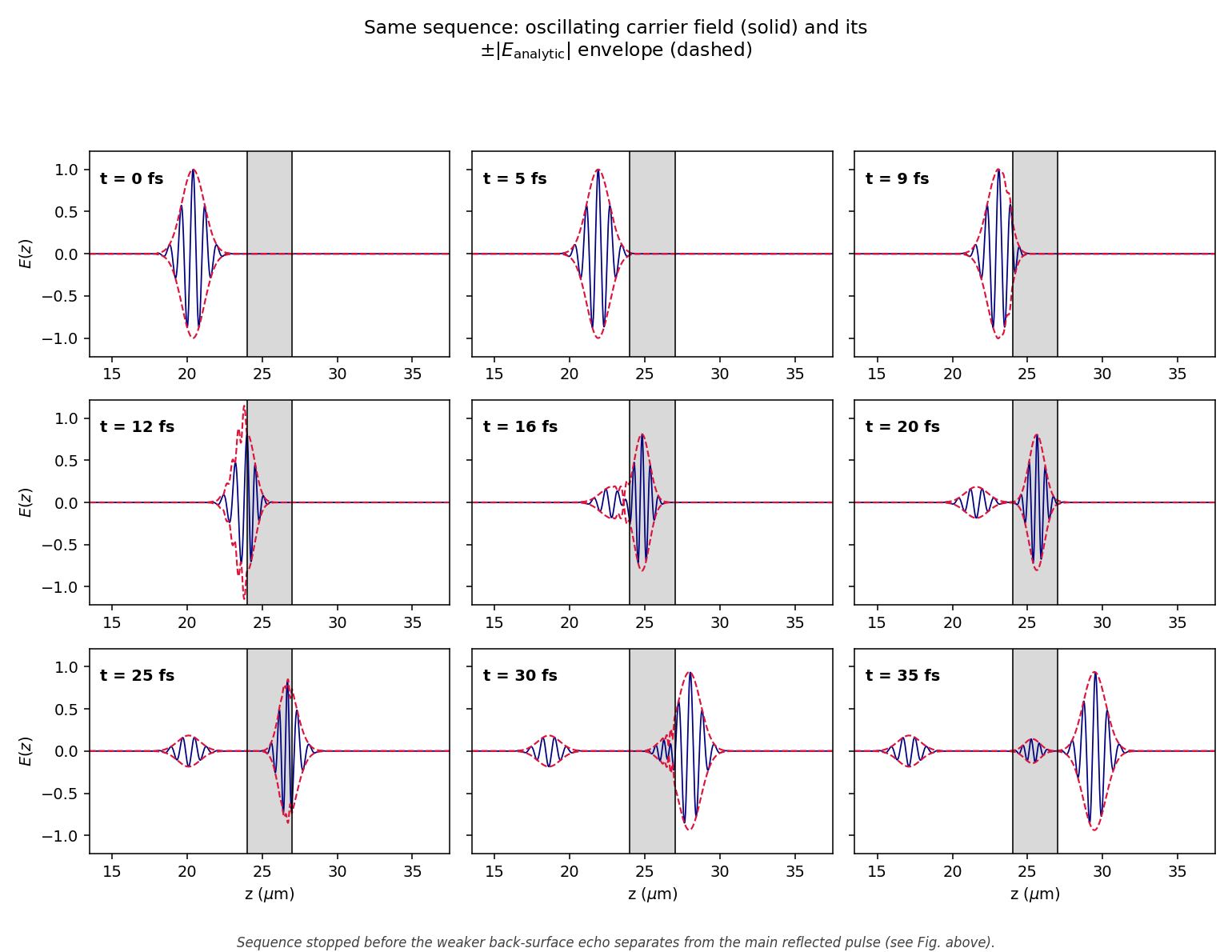}
\caption{FDTD--ADE: the same nine instants, oscillating carrier field $E(z,t)$ (solid) with its $\pm|E_{\mathrm{analytic}}(z,t)|$ envelope (dashed).}
\label{fig:movie_carrier_fdtd}
\end{figure}

\begin{figure}[!htbp]
\centering
\maybeincludegraphics[width=0.95\textwidth]{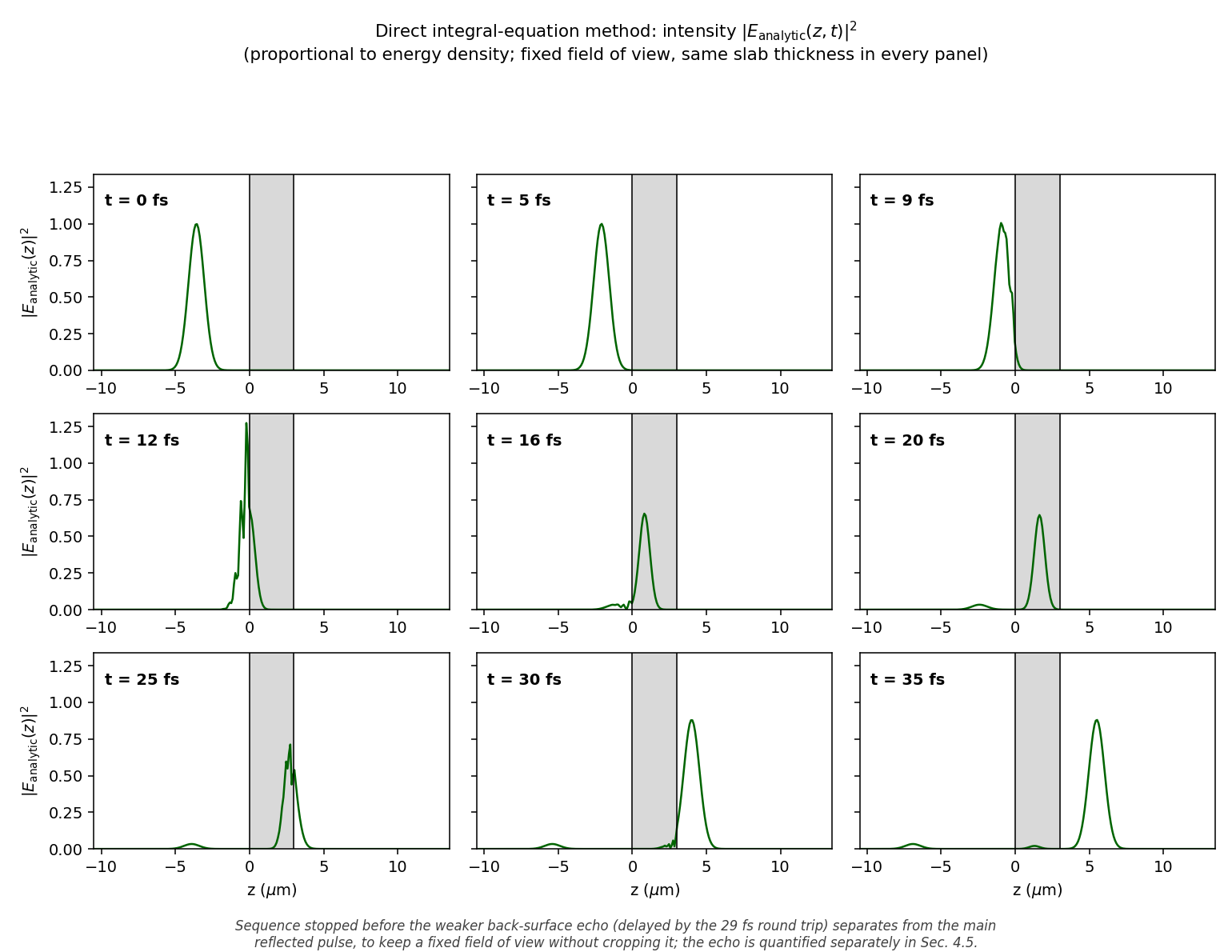}
\caption{Direct integral-equation method: intensity $|E_{\mathrm{analytic}}(z,t)|^2$, same nine snapshot times, obtained from Eq.~\eqref{eq:external} as a post-processing step on the internally-computed polarization history -- no additional time-marching outside the slab.}
\label{fig:movie_intensity_direct}
\end{figure}

\begin{figure}[!htbp]
\centering
\maybeincludegraphics[width=0.95\textwidth]{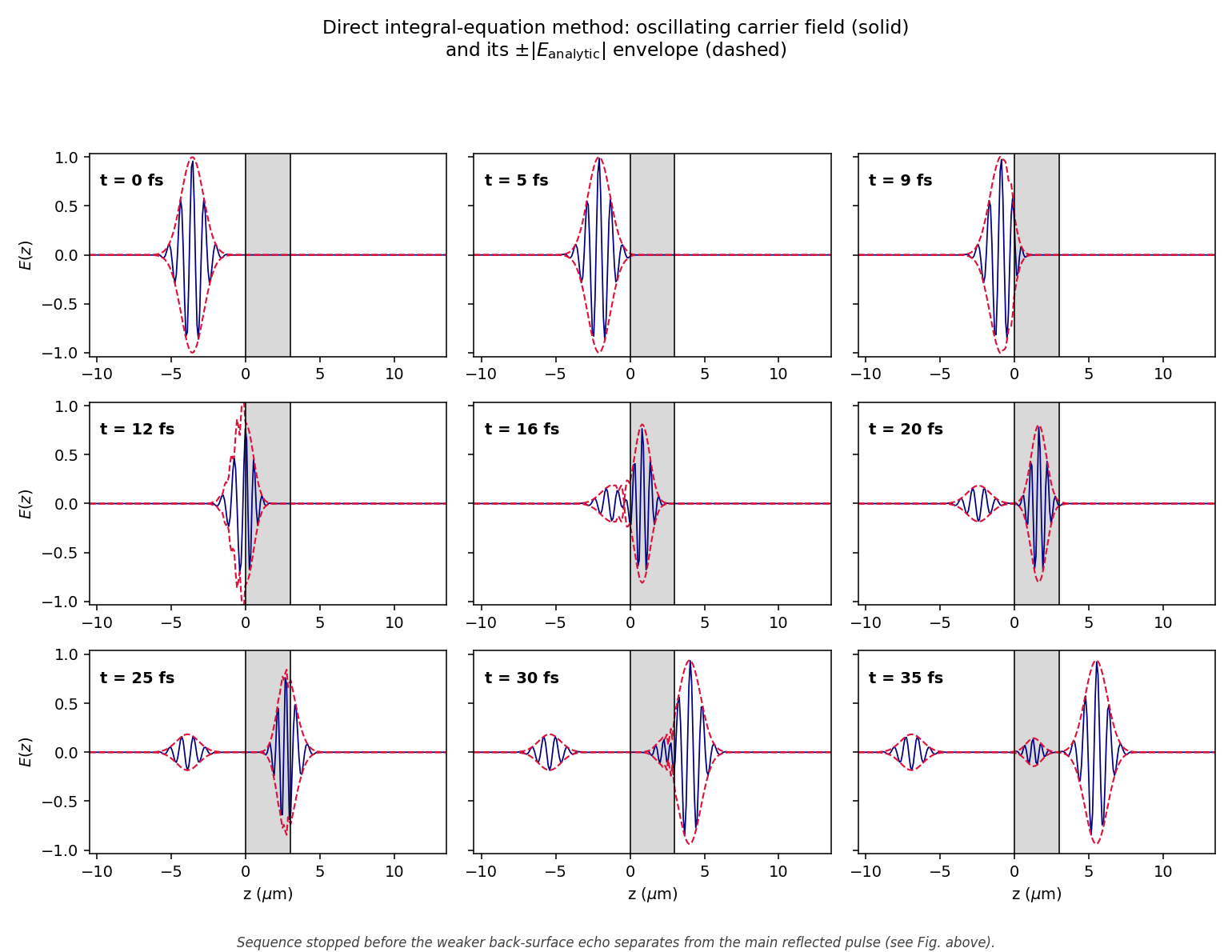}
\caption{Direct integral-equation method: oscillating carrier field (solid) and envelope (dashed), same nine instants.}
\label{fig:movie_carrier_direct}
\end{figure}

Sections~\ref{sec:methodA}--\ref{sec:results} compare the two methods through integrated observables and through the field in a fixed, narrow spatial window around the thin slab. This section adds a different, deliberately pedagogical view of the same equivalence: nine-frame, movie-style snapshots of a single representative case (the thick slab, $L=3\,\mu\mathrm{m}$) as the pulse approaches, interacts with, and separates from the slab, in the spirit of Goldberg, Schey \& Schwartz's classic wave-packet movies \cite{GSS1967}. As in their own figures, a single, fixed spatial window, centered on the slab, is used for all nine panels, so the slab keeps the same apparent thickness throughout the sequence rather than changing size from frame to frame. This choice comes at a cost: the sequence is stopped once the reflected and transmitted wavepackets have clearly separated but before the weaker back-surface echo, delayed by the $29.0\,\mathrm{fs}$ round trip, would need extra room to stay in view; the echo is not a loss of physics, since it is already quantified and discussed in Sec.~\ref{sec:results} (Table~\ref{tab:energy}) -- it is simply left out of this particular figure for legibility, exactly as Goldberg, Schey \& Schwartz themselves photograph only the central portion of their own domain for the same reason \cite{GSS1967}. Table~\ref{tab:params} applies unchanged. \rev{Figures~\ref{fig:movie_intensity_fdtd} and~\ref{fig:movie_carrier_fdtd} show the FDTD--ADE result, Figs.~\ref{fig:movie_intensity_direct} and~\ref{fig:movie_carrier_direct} the direct integral-equation result, at the same nine instants.}

Frames $t=9$--$16\,\mathrm{fs}$ (using the local time labeling of this section, distinct from the absolute clock of Secs.~\ref{sec:methodA}--\ref{sec:results}) show a secondary lobe separating from the main pulse, with peak amplitude approaching the Fresnel amplitude reflection coefficient $r=(n-1)/(n+1)=0.184$: the front-surface reflected wavelet, resolved as a separate feature only a few femtoseconds after the interaction begins. The last three frames show this reflected wavelet and the transmitted pulse clearly separated and moving apart at $c$, tracked directly from their peak positions across frames; the back-surface echo is not shown, for the reason stated above.

The two methods reproduce this sequence feature for feature: the same interference fringes between the still-arriving incident tail and the newly-forming reflected wavelet during the interaction, and the same clean separation of the two visible wavepackets in the later frames. Reading the reflected and transmitted fields directly off probes in this simulation geometry, over the full simulation time (not only the displayed snapshot window), gives $R=0.0625$, $T=0.8826$ (Method A) and $R=0.0618$, $T=0.8840$ (Method B) -- consistent, to the precision reported, with the values of Table~\ref{tab:energy} obtained from the independent simulation geometry of Secs.~\ref{sec:methodA}--\ref{sec:results}, as expected since both geometries solve the identical physical problem.

\section{Discussion}
\label{sec:discussion}

The two independently discretized formulations reproduce the same principal field behavior and the same integrated observables to within the numerical differences reported above. The agreement is especially close in the thin-slab case, which is the more demanding configuration because the multiple reflected contributions overlap within the pulse duration. It is corroborated at a third, independent level in Sec.~\ref{sec:movie}, where a differently-shaped, differently-windowed simulation domain, run with each method's own solver, reproduces the same $R$ and $T$ to within the same discretization tolerance -- an end-to-end check that the agreement documented in Secs.~\ref{sec:methodA}--\ref{sec:results} is not an artifact of one particular choice of domain or probe placement.

The comparison should, however, be interpreted with the numerical limitations made explicit. First, the two methods do not use identical numerical discretizations: Method A uses $S=0.4$ for its FDTD time step, whereas Method B uses $S=0.5$ to align the retarded delays with the time grid. Second, while the absorbed fraction is now independently corroborated (Table~\ref{tab:energy}) rather than only inferred from $1-R-T$, this required a dedicated volume integral of the oscillator's dissipation rate, distinct from the field-only quantities $R$ and $T$; it is worth flagging explicitly, since an inferred balance alone would not have constituted an independent test. Third, the absence of temporal interpolation in the internal coupling of Method B does not remove spatial quadrature or time-discretization error, nor the temporal interpolation used in Eq.~\eqref{eq:external} for points outside the slab. Finally, \rev{the comparison with the exact values of Sec.~\ref{sec:exact} shows that, at the resolutions used, Method~B is the closer of the two to the exact thick-slab result. This reflects the coarser grid of the FDTD run rather than an intrinsic superiority of the integral formulation: both schemes are second-order accurate, and both converge to the same exact values.}

A further methodological note concerns the movie-style visualization of Sec.~\ref{sec:movie}. An earlier version of the post-processing pipeline used to build Figs.~\ref{fig:movie_intensity_direct}--\ref{fig:movie_carrier_direct} computed the field envelope via a Hilbert transform along the \emph{spatial} axis at each fixed snapshot time; this does not correctly represent the interference between the two counter-propagating wave components (incident tail and newly-formed reflection) present during the interaction, and visibly smoothed away fringe structure that Method A resolves in that regime. The corrected pipeline, used throughout this report, applies the Hilbert transform along \emph{time} at each spatial point, matching how the envelope is computed in Secs.~\ref{sec:results}--\ref{sec:movie} for Method A. We independently verified that the space-time and envelope maps of Figs.~\ref{fig:3}--\ref{fig:4}, generated by a separate script, were not affected by this issue. The episode is a reminder that agreement on integrated quantities alone does not guarantee that every diagnostic built on top of a simulation is itself correct, and that comparing two independent methods field-for-field, not only number-for-number, is a genuinely more stringent test.

Within these limits, the results provide direct numerical evidence that the retarded multiple-scattering integral equations and the local Maxwell--Lorentz equations lead to the same physical field for the tested few-cycle pulse and slab configurations. The computational trade-off is clear: the local FDTD--ADE method is much faster, whereas the direct integral formulation provides an independent and physically transparent representation of the same retarded interactions, and -- once the self-interaction and retarded-lookup issues of Sec.~\ref{sec:methodB} are resolved -- computes the field outside the slab as inexpensive post-processing rather than additional time-marching, which made it convenient for the wide-field-of-view visualization of Sec.~\ref{sec:movie}.

\section{Conclusions}
\label{sec:conclusions}

We have solved the retarded integral equations directly in the time domain using a non-iterative treatment of the zero-delay self-interaction and exact integer-step retrieval of the retarded history. The resulting fields and integrated observables closely reproduce those obtained from the independently discretized FDTD--ADE formulation, across three independent checks: a narrow-window field and spectral comparison, an energy balance corroborated by an independently-computed absorbed fraction, and a wide-field-of-view, movie-style visualization run on a differently-shaped simulation domain. Agreement persists in the sub-wavelength thin-slab regime, where the naive Neumann/Picard iteration is not a practical convergent approach for the parameters considered.

The numerical evidence therefore supports the analytic equivalence established in \rev{Paper~I} and shows that it survives an independent time-domain implementation, at the level of the fields, the principal observables, and a genuinely independent energy-conservation check\rev{, and both numerical solutions converge to an exact reference for the integrated energy fractions}. Remaining directions for a still more stringent validation include a systematic refinement study for both methods, quantitative field and spectral error norms beyond the single RMS value reported here, and a dedicated asymptotic or spectral analysis if a rigorous identification of Sommerfeld--Brillouin precursor structure is sought.

As anticipated in Sec.~\rev{\ref{sec:intro}}, the two formulations cross-validated above instantiate, for a
linear, classical, one-dimensional problem, a duality with a well-known historical
counterpart. Schwinger~\cite{Schwinger} and Tomonaga~\cite{Tom} built quantum electrodynamics from local field operators
obeying canonical equations of motion --- in the present classical problem, the same
computational character as Method~A, which advances the field and the polarization forward in
time through a local, auxiliary differential equation. Feynman~\cite{Fey} instead wrote the solution
directly as a sum over propagators\rev{ (his causal propagator, which carries positive frequencies forward in time; in the present classical, linear problem its role is taken by the retarded Green's function)}, the same structure as Eq.~\rev{\eqref{eq:external}}, where the field
outside the slab is obtained as an integral of the source polarization against the retarded
Green's function of the wave operator, rather than as the outcome of a local time march.
Dyson~\cite{Dyson} proved that the two quantum-electrodynamical formulations, despite their entirely
different computational apparatus, are mathematically equivalent. The agreement documented
above between Method~A and Method~B --- established independently and cross-checked at the
level of the fields, the integrated observables, and the energy balance --- is, at an
incomparably more modest scale, the same statement verified numerically rather than proved
analytically: a local, differential description and a non-local, retarded-propagator
description of one causal linear system are dual representations of the same physics, not two
competing claims to be adjudicated against each other.

\section*{Acknowledgments}
The author thanks Antoniangelo Agnesi, former student and colleague, for fruitful discussions and suggestions during the preparation of this work. \rev{The author also acknowledges the use of Claude (Anthropic), an AI assistant, for help with the numerical analysis and the editorial revision; the author takes full responsibility for the content.}

\appendix
\section{Glossary of numerical-method terms}
\label{app:glossary}

\paragraph{Courant stability limit.}
For the standard one-dimensional vacuum leapfrog wave scheme used in Method A, the Courant number is
\begin{equation}
S\equiv\frac{c\Delta t}{\Delta z}.
\end{equation}
The usual stability condition for this one-dimensional scheme is $S\le1$. Here Method A uses $S=0.4$. Method B uses $S=0.5$ not primarily as a stability choice, but to make every spatial retardation an integer number of time steps.

\paragraph{Retarded time and retarded integral equation.}
In the non-local formulation, the field at $(z_i,t_n)$ depends on source values at the retarded time $t_n-|z_i-z_j|/c$. This accounts for finite propagation speed and requires access to the source history. The numerical treatment of that history is the central computational issue addressed by Method B \cite{Davies1994,Chapko2019}.

\paragraph{Zero-delay self-interaction.}
When $z_j=z_i$, the retardation vanishes. The resulting current-time contribution makes the discrete field update implicit. In Method B this term is isolated and eliminated algebraically. The term is called ``self-interaction'' here; the adjective ``singular'' is avoided because the present one-dimensional discretized kernel is not itself being claimed to possess a mathematical singularity at $z_j=z_i$.

\paragraph{Exact retarded lookup.}
For $S=0.5$, the delay satisfies $|z_i-z_j|/c=2|i-j|\Delta t$. The required source value therefore lies exactly on the stored time grid. ``Exact'' here means exact with respect to temporal interpolation; it does not imply an absence of other discretization errors. Points outside $[0,L]$ (Sec.~\ref{sec:movie}) do not benefit from this alignment and use linear interpolation instead (Eq.~\eqref{eq:external}).

\paragraph{Richardson (empirical) order of convergence.}
For three successively halved grid spacings, an observed order can be estimated from
\begin{equation}
p\approx\log_2\left|\frac{f(\Delta z)-f(\Delta z/2)}{f(\Delta z/2)-f(\Delta z/4)}\right|.
\end{equation}
This is an empirical diagnostic and should only be interpreted as an asymptotic convergence order when the calculation is demonstrably in the asymptotic regime.

\paragraph{Mur absorbing boundary condition.}
A finite computational domain requires approximate absorbing boundaries to reduce spurious reflections. The first-order Mur condition used in Method A is a one-way-wave approximation; in the present strictly one-dimensional geometry it is appropriate for the intended outgoing-wave boundary treatment \cite{Mur}.

\paragraph{Wall time (CPU time).}
Wall time is the elapsed real time required to execute a complete simulation. The FDTD--ADE update is local and scales approximately as $O(N_z)$ per time step, whereas the straightforward retarded summation scales as $O(N_z^2)$ per time step in Method B.

\paragraph{Group velocity versus phase velocity.}
In a dispersive medium, the phase velocity at the carrier is conventionally written $v_p=c/n$, while the group velocity near that frequency is $v_g=c/n_g$. The latter describes the motion of a narrowband pulse envelope under the usual assumptions of weak distortion. For the few-cycle pulse considered here, the comparison with $n_gL/c$ is therefore best regarded as a consistency check rather than as a universal statement about the velocity of energy or information.

\section{Independent energy-conservation check}
\label{app:energy}

The absorbed fraction quoted as $A_{\mathrm{indep}}$ in Table~\ref{tab:energy} is obtained, for each method separately, from Poynting's theorem applied to the coupled Maxwell--Lorentz system, without reference to $R$ or $T$. Writing the Lorentz oscillator equation as $\ddot P+2\gamma\dot P+\omega_0^2P=\kappa E$ and combining it with the electromagnetic Poynting balance gives a total (field-plus-oscillator) energy density obeying
\begin{equation}
\frac{\partial}{\partial t}\big(u_{EM}+u_{\mathrm{osc}}\big) + \frac{\partial S}{\partial z} = -\frac{2\gamma}{\kappa}\left(\frac{\partial P}{\partial t}\right)^2 \le 0,
\label{eq:dissipation}
\end{equation}
where $u_{\mathrm{osc}}=(2\kappa)^{-1}(\dot P^2+\omega_0^2P^2)$. Integrating the right-hand side of Eq.~\eqref{eq:dissipation} over the slab volume and over the full simulation time, and normalizing by the incident pulse energy, gives $A_{\mathrm{indep}}$. In the discrete implementation this is
\begin{equation}
A_{\mathrm{indep}} = \frac{1}{E_{\mathrm{inc}}}\sum_n \Delta t \sum_i \Delta z\,\frac{2\gamma}{\kappa}\,\dot P_i^2(t_n),
\label{eq:Aindep-discrete}
\end{equation}
evaluated with the centered-difference estimate of $\dot P_i(t_n)$ already used in Eq.~\eqref{eq:Pdot}, over the slab grid in each method. That this integral, computed from the polarization dynamics alone, agrees with $1-R-T$ (computed from the boundary fields alone) to the precision shown in Table~\ref{tab:energy} -- three significant figures where the underlying run supports it -- in both methods and both slab thicknesses is the genuinely independent confirmation referenced in Sec.~\ref{sec:discussion}.

\end{document}